\documentstyle[psbox,psfig]{WORKSHOP}

\def\deg{$^\circ$}

\def\fdeg{\hbox{$.\!\!^{\circ}$}}

\def\0142{4U~0142+61}
\def\axp1048{1E~1048.1-5937}

\def\rxs1708{1RXS~J1708-40}
\def\xte1810{XTE~J1810-197}
\def\1841{1E~1841-045}
\def\ax1845{AX~J1845.0-0258}
\def\2259{1E~2259+586}

\begin{document}

% TITLE OF THE PAPER
%  If the title is too long for a single line, you can split it 
%  by putting two backslashes. 
%  You might want to put the subtitle. Then it should be inserted 
%  within {\large\sf  }.
%  e.g.:  
%     \title{ Too Long Title \\ for one line \\
%     {\large\sf Subtitle} }
\title{
Hard X-ray/soft $\gamma$-ray Characteristics of the Persistent Emission from Magnetars \\ 
{\large\sf  -- Results based on multi-year INTEGRAL, RXTE and XMM Newton observations --} % SUBTITLE
}

% AUTHOR(S) 
\author{
L. Kuiper,$^1$ P.R. den Hartog$^1$ and W. Hermsen$^{1,2}$ \\[12pt]  % TO BE SPACED WITH ONE LINE
%
% INSTITUTES OF AUTHORS
$^1$ SRON, Netherlands Instituut for Space Research, Sorbonnelaan 2, 3584 CA, Utrecht, The Netherlands  \\
$^2$ Sterrenkundig Instituut ``Anton Pannekoek'', University of Amsterdam, Kruislaan 403, 1098 SJ, Amsterdam, \\
The Netherlands \\
%
% please put the first author's initial and e-mail address below
{\it E-mail(LK): L.M.Kuiper@sron.nl} 
%            \_ Initial      \
%                             \_ E-mail address
}

\abst{In this paper the current status of high-energy research on the hard X-ray characteristics of the persistent 
emission from magnetars is reviewed. Focus is put on
recent intriguing results for \rxs1708\, from phase resolved spectral analysis over a 
2 decades wide energy band ($\sim 3-300$ keV) combining contemporaneous RXTE, XMM and
INTEGRAL data. For \1841 and SGR 1806-10 we also present updated results. The perspective for future MAXI observations for this source class is also 
addressed.
}

\kword{Anomalous X-ray pulsars --- Soft gamma-ray repeaters --- hard X-rays --- soft gamma-rays --- persistent emission}

\maketitle
\thispagestyle{empty}

\section{Introduction}
Anomalous X-ray pulsars (AXP) belong to a class of rare objects
concentrated along the galactic plane and emitting pulsed X-rays with
pulse periods in the range $\sim 2-12$ s, characteristic spin-down
time scales of $\sim 10^3-10^5$ year and surface magnetic dipole 
field strengths well above the quantum critical value of $B_{QED}\simeq 4.4\times 10^{13}$ Gauss. 
The fact that the observed X-ray
luminosity is much larger than the spin-down power excludes an
interpretation in which the (pulsed) X-ray emission originates from a
spin-down powered pulsar (see e.g. Kaspi 2007 or Woods \& Thompson
2006 for reviews on AXPs). On the other hand the lack of apparent bright
optical counterparts and the absence of periodic modulation of the X-ray
pulsations exclude an interpretation involving accreting compact sources.

Currently, models based on the decay of
very strong magnetic fields ($10^{14}-10^{15}$ G) - so called
``magnetar'' models (e.g. Thompson, Lyutikov \& Kulkarni 2002 and
Thompson \& Duncan 1996 and references therein) - explain the
observed soft X-ray characteristics (0.5--10 keV) of AXPs at a
satisfactory level. Bursting and (rotational) glitching behaviour have been 
observed from AXPs.  These phenomena are common for Soft Gamma-ray
Repeaters (SGRs) for which the magnetar model originally was
developed.  These shared properties provide strong evidence that both AXPs
and SGRs are members of the same source class.

The X-ray spectra of AXPs in the 0.5--10 keV band are very soft and
can be described empirically by a black body plus a power-law model. The softness
of the spectra below $\sim$10 keV (power-law indices all $>$2)
predicted non-detections for energies above $\sim$10 keV and thus
explains the initial ignorance of studying the spectral properties of
AXPs at energies above 10 keV.

It was, therefore, a big surprise that INTEGRAL IBIS-ISGRI (20-300 keV) measured 
hard X-rays from the direction of three AXPs. 

Molkov et al. (2004) reported the discovery of an INTEGRAL source at the position of AXP \1841\,
in SuperNova Remnant (SNR) Kes 73 for energies up to 120 keV.  This was followed up 
by Kuiper et al. (2004), who analysed archival RXTE PCA and HEXTE data from
monitoring observations, to prove that the hard
X-ray emission comes from the AXP and not from the SNR. They
discovered pulsed hard X-ray/soft $\gamma$-ray emission
up to $\sim$150 keV.

Since then Revnivtsev et al. (2004) and den Hartog et al. (2006)
published the INTEGRAL detections of AXP \rxs1708 and AXP \0142, respectively. These
works were again followed up by a search in archival RXTE PCA and HEXTE
data (Kuiper et al. 2006) and pulsed hard X-ray emission was detected.

Currently ten AXPs (one should be considered as a candidate AXP) and five SGRs are known. 
For three persistently ``bright" AXPs hard X-ray spectra ($>$20 keV) 
have been measured with INTEGRAL up to at least 150 keV. For the other seven AXPs 
no persistent hard X-ray detections can yet be claimed from the INTEGRAL 
data\footnote{Pulsed high-energy emission up to $\sim 24$ keV has been detected from 1E 2259+586
in RXTE PCA data combining $\sim 750$ ks of exposure.}.
Also, from two SGRs, SGR 1806-10 (Mereghetti et al. (2005), Molkov et al. (2005)) and SGR 1900+14
(G\"otz et al. 2006), persistent hard X-ray emission has been detected.
In the magnetar model persistent hard X-ray emission was initially ignored.
Since the new findings described in this review, several theoretical attempts to explain the
hard X-ray characteristics have been made (Heyl \& Hernquist 2005(a,b), 2007; Thompson \& Beloborodov 2005; Beloborodov \& Thompson 2007; Baring \& Harding 2007, 2008). However, so far theoretical consensus has not
been reached and the observed phenomenology remains unexplained.
In this paper the observational aspects are addressed and we summarize the hard X-ray/soft $\gamma$-ray emission characteristics above $\sim 20$ keV for the currently known sample of five magnetars emitting hard X-rays.

\begin{figure}[t]
\centering
\psbox[xsize=8cm]
{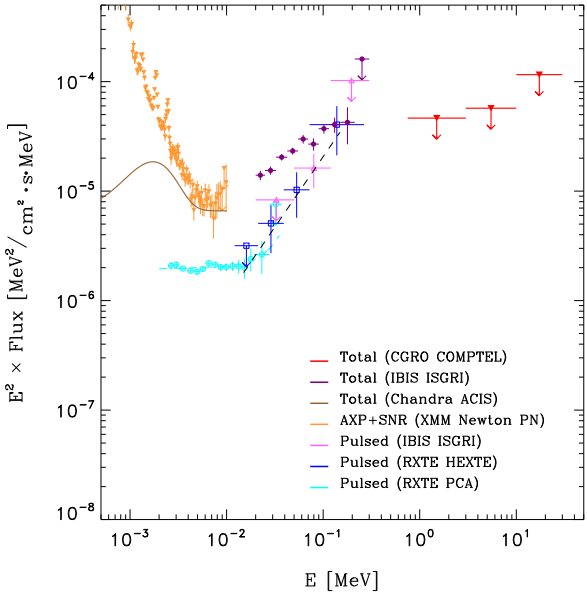}
\caption{High-energy total and pulsed spectra of \1841\, combining RXTE PCA, HEXTE and INTEGRAL ISGRI data for the $\sim 3-300$ keV pulsed spectrum; XMM Newton EPIC and INTEGRAL ISGRI for the total spectrum from \1841\, and Kes 73; and Chandra ACIS for the total (=pulsed plus DC) emission from \1841. Also, CGRO COMPTEL $2\sigma$ upper-limits are shown. Note, the drastic upturn of the pulsed emission near 15 keV. The emission is 100\% pulsed beyond $\sim 100$ keV.}
\label{spectrum_1841}
\end{figure}

%%%%%%%%%%%%%%%%%%%%%%%%%%%%%%%%%%%%%%%%%%%%%%%%%%%%%%%%%%%%%%%%%%%%%%%%%%%%%%%%%%%%%%%%%%%%%%%%
\section{Persistent hard X-ray/soft $\gamma$-ray magnetars}
%%%%%%%%%%%%%%%%%%%%%%%%%%%%%%%%%%%%%%%%%%%%%%%%%%%%%%%%%%%%%%%%%%%%%%%%%%%%%%%%%%%%%%%%%%%%%%%%

\subsection{\1841\, (in SNR Kes 73)}

\1841\, is located at the very center of SNR Kes 73. In the source catalogue published by
Molkov et al. (2004), surveying the Sagittarius Arm tangent region, a hard X-ray (18-120 keV) source was reported from imaging studies positionally consistent with \1841\, and Kes 73.
Subsequent work by Kuiper et al. (2004) showed that the emission above $\sim 20$ keV could be
attributed, surprisingly, to the anomalous X-ray pulsar because the pulsed fingerprint was detectable up to 
$\sim 150$ keV using RXTE HEXTE (15-250 keV) data. The pulsed spectrum above $\sim 10$ keV appeared to be very hard with a photon index of $0.94\pm0.16$. The spectral picture for energies
between 0.5 and 300 keV was updated by Kuiper et al. (2006) including now for the first time 
the total flux measurements from INTEGRAL IBIS ISGRI derived for seven logarithmically binned energy bands between 20 and 300 keV using INTEGRAL data collected between March 10, 2003 and Nov. 8, 2004. The RXTE HEXTE total flux 
values were abandoned in that work because it was realized that these were inaccurate due to
the non-imaging nature of the instrument (on-off observation strategy with relatively large field of views).
The total emission from \1841\, above 20 keV was consistent with a power-law model with a photon index of $1.32\pm 0.11$. Comparison of the pulsed and total spectra above $\sim 20$ keV indicated that the emission is consistent with being 100\% pulsed for energies beyond 100 keV.
Assuming that the emission level above 20 keV is reasonably stable CGRO COMPTEL measurements performed earlier between 1991-2000 in the 0.75-30 MeV band demand a spectral break in the range 300 - 750 keV.
Also, Kuiper et al. (2006) confirmed the presence of the pulsed signal in the $>20$ keV INTEGRAL ISGRI data.
In this work an updated INTEGRAL ISGRI total spectrum (20-300 keV) in ten logarithmically binned energy bands is presented using data from INTEGRAL observations performed between March 10, 2003 and May 26, 2006. Spectral analysis adopting a power-law model yields a photon index of $1.39\pm 0.05$ and a 20-100 keV flux of $(5.9\pm 0.2)\times 10^{-11}$ erg/cm$^2$s. The 20-300 keV spectral data do still not reveal a spectral break in this range. The current high-energy picture is shown in Fig.\ref{spectrum_1841}.
%%%%%%%%%%%%%%%%%%%%%%%%%%%%%%%%%%%%%%%%%%%%%%%%%%%%%%%%%%%%%%%%%%%%%%%%%%%%%%%%%%%%%%%%%%%%%%%%
\begin{figure}[h]
\vspace{-2mm}
\centering
\centerline{\psfig{figure=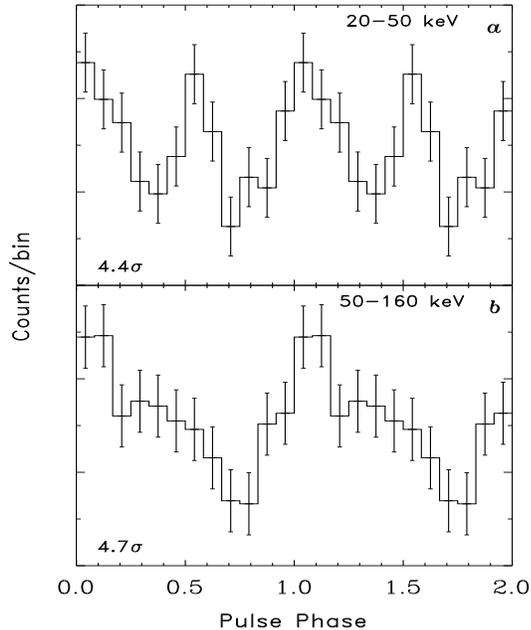,bbllx=145pt,bblly=195pt,bburx=420pt,bbury=610pt,width=7cm,height=8.5cm}}
\caption{The hard X-ray/soft $\gamma$-ray pulse profiles of \0142\ as measured by INTEGRAL IBIS ISGRI for the 20-50 (top) and 50-160 (bottom) keV bands. Significances of $4.4\sigma$ and $4.7\sigma$ are obtained for both bands, respectively.}
\label{isgri_4u_pp}
\end{figure}

\subsection{\0142}
The detection of \0142\, at hard X-rays was announced by den Hartog et al. (2004) using INTEGRAL ISGRI data from survey observations of the Cassiopeia region performed in December 2003. More detailed information was provided by den Hartog et al. (2006), particularly, they measured a very
hard, $\Gamma=0.73\pm0.17$, power-law like spectrum for the total emission over the 20-150 keV range, and a luminosity in the 20-100 keV band of about 440 times the rotational energy loss. Moreover, they reported CGRO COMPTEL upper limits for the 0.75-30 MeV band demanding a
spectral break between $\sim 100$ and $750$ keV, assuming stable high-energy emission during the different CGRO and INTEGRAL observation epoches. 

\begin{figure}[t]
\centering
\psfig{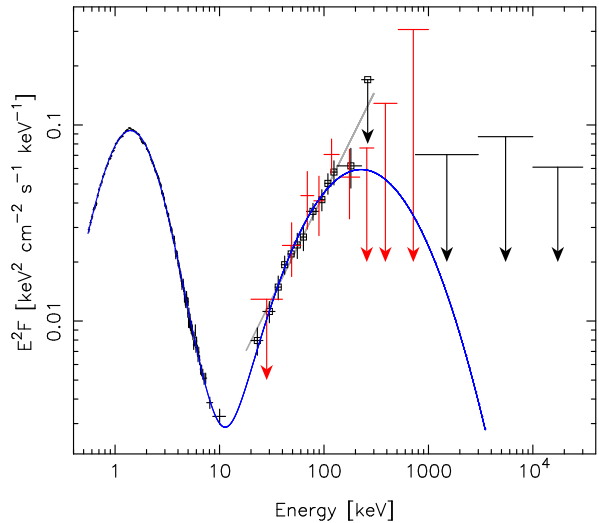}
\caption{The total high-energy spectrum of \0142 from 0.5 keV up to 30,000 keV combining contemporaneous INTEGRAL ISGRI and SPI (red) data along with XMM Newton EPIC data. The best fit model (blue; a double log-parabola spectral model over the 0.5-1,000 keV band) has also been superposed
together with CGRO COMPTEL upper limits ($2\sigma$). A spectral break is evident from these spectral measurements.}
\label{he_spec_4u}
\end{figure}

Revisiting archival RXTE data collected over the period March 28, 1996 - Sept. 18, 2003 Kuiper et al. (2006) found the pulsed signal of \0142\, up to $\sim 50$ keV in RXTE HEXTE data. The pulsed spectrum over the 2-32 keV range could be described adequately by a double power law model with a soft and hard index of $\Gamma_s=4.09\pm0.04$ and $\Gamma_h=-0.8\pm0.10$, respectively. Such a drastic hardening, around 10 keV,
had never been observed for any source.

den Hartog et al. (2008a) published the most detailed findings at high-energies for \0142. By combining 4.5 Ms of
screened INTEGRAL ISGRI data from observations performed between Dec. 12, 2003 and Aug. 13, 2006 they detected
pulsed emission from \0142 up to $\sim 160$ keV. The two profiles for the 20-50 keV ($4.4\sigma$) and 50-160 keV ($4.7\sigma$) bands are shown in Fig.\ref{isgri_4u_pp}(a,b), respectively. The integral 20-160 keV pulsed signal reached a $6.5\sigma$ significance.
The total high-energy signal ($>$ 20 keV) was studied for different epoches and no evidence was found for significant long-term time variability. Both the flux and photon index were stable within the 17\% level ($1\sigma$).
These authors also reported evidence for a spectral break in the total spectrum above 20 keV combining contemporaneous INTEGRAL ISGRI (20-300 keV) and SPI (20-1000 keV) data: the $\nu F_{\nu}$ spectrum peaks at
$228_{-41}^{+65}$ keV. The total spectrum from 0.5 to 30,000 keV is shown in Fig. \ref{he_spec_4u} 
displaying XMM Newton, INTEGRAL ISGRI and SPI and CGRO COMPTEL measured flux values and upper limits.
The best fit model (blue; see for details den Hartog et al., 2008a) is superposed.

In den Hartog et al. (2008a) phase-resolved spectral analysis results were presented for three relatively broad phase intervals with boundaries dictated by the observed emission features in the underlying pulse-profiles
over the 0.5 to 300 keV range. Data from RXTE PCA, INTEGRAL ISGRI, XMM Newton and (non-contemporaneous) ASCA GIS
were combined for this purpose using consistent phase aligning. 
The ASCA GIS data showed different pulse morphologies and flux levels compared to data from the later performed XMM Newton and RXTE PCA observations. This might be related to a glitch happening near the ASCA observation. 
Clear trends are visible in the phase-resolved pulse-emission spectra (see for more details den Hartog et al., 2008a).

%%%%%%%%%%%%%%%%%%%%%%%%%%%%%%%%%%%%%%%%%%%%%%%%%%%%%%%%%%%%%%%%%%%%%%%%%%%%%%%%%%%%%%%%%%%%%%%%

\subsection{\rxs1708}
High-energy emission ($>20$ keV) from \rxs1708 was reported by Revnitsev et al. (2004) and was confirmed by 
Kuiper et al. (2006) using more INTEGRAL data. Revisiting archival RXTE data
(Jan. 12, 1998 - Oct. 26, 2003) and using INTEGRAL ISGRI data these authors detected pulsed emission even
above 75 keV.

\begin{figure}[t]
\centering
\psfig{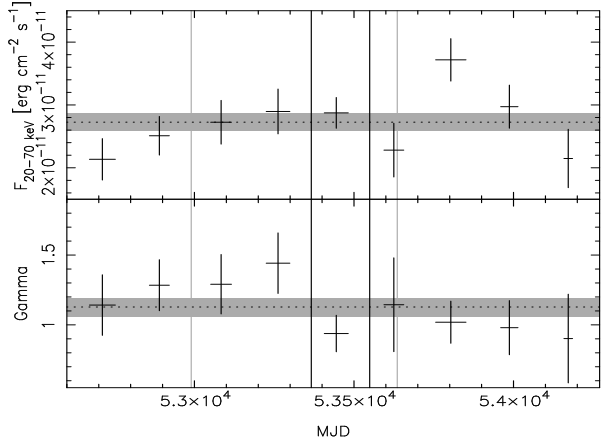}
\caption{Temporal behaviour of the hard X-ray flux in the 20-70 keV band (top) and photon index (bottom) of \rxs1708\, as measured by INTEGRAL IBIS ISGRI. No significant flux variations have been detected. Epoches of glitches and possible glitches
are indicated by black and grey vertical lines, respectively.}
\label{rxs_flux_var}
\end{figure}

\begin{figure}[t]
\centering
\psfig{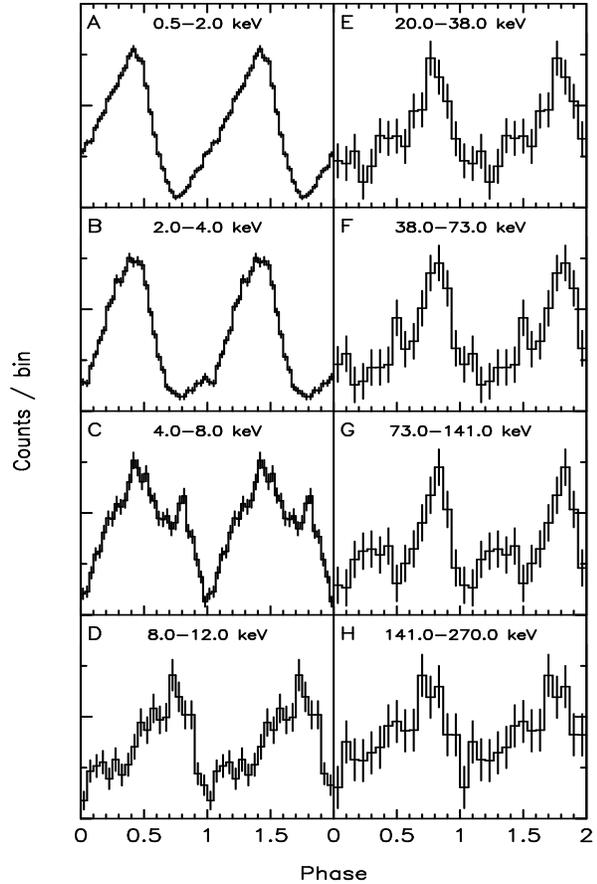}
\caption{Pulse profiles of \rxs1708 from soft to hard X-rays using XMM Newton EPIC PN data (0.5-12 keV) and INTEGRAL ISGRI
data (20-270 keV). Note the stable morphology for energies beyond 20 keV.}
\label{rxs_pp}
\end{figure}

den Hartog et al. (2008b) presented new updates on the high-energy emission research using now all
available (Jan. 29, 2003 -- Oct. 5, 2006) INTEGRAL ISGRI data with \rxs1708 in the field of view.
The total screened good time exposure was $\sim 12$ Ms, which translated in an effective on-axis exposure
of $\sim 5.2$ Ms.
The long-term variability in the 20-300 keV band was studied by performing spectral analyses on a half-year and
yearly base. No significant variability was found either in flux or in photon index (see Fig. \ref{rxs_flux_var}) in contradiction with G\"otz et al. (2007), who claimed long-term hard X-ray variability in relation to glitch activity.
Therefore, den Hartog et al. (2008b) considered the full ISGRI dataset and derived a time-averaged total spectrum 
(20-300 keV), described properly by a power-law model with photon index $\Gamma = 1.13\pm 0.06$ and a 20-150 keV flux of 
$(6.61\pm0.23)\times 10^{-11}$ erg/cm$^2$s. No hint for a spectral break could be found below 300 keV. Assuming stable 
high-energy emission from 1991 up to 2006 the CGRO COMPTEL $2\sigma$ upper limits require a spectral break somewhere 
between 300 and 750 keV.

\begin{figure}[t]
\centering
\psfig{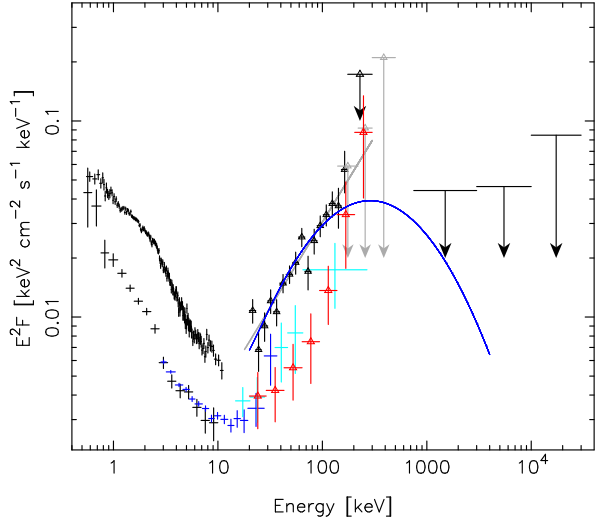}
\caption{The total and total pulsed high-energy (0.5-300 keV) spectrum of \rxs1708\, as measured by XMM Newton (total, pulsed), RXTE PCA (pulsed), RXTE HEXTE (pulsed) and INTEGRAL IBIS ISGRI (total, pulsed). CGRO COMPTEL (0.75-30 MeV) $2\sigma$ upper limits are included for the total emission as well (see for more details den Hartog, 2008b).}
\label{rxs_he_spc}
\end{figure}

A timing analysis of the full ISGRI dataset yielded a considerably improved pulse profile, now, with a detection 
significance of $12.3\sigma$ for the 20-270 keV range. Pulsed emission is detected up to $\sim 270$ keV. The morphology 
of the pulse profiles as a function of energy from 0.5 up to 270 keV is shown in Fig. \ref{rxs_pp}. XMM-Newton EPIC PN data have 
been used for the part below 12 keV. Drastic morphology changes are visible in this collage.
From these pulse profiles the total pulsed spectrum of \rxs1708 has be derived. Both the total and total pulsed spectrum
(RXTE PCA and HEXTE data are also included) of \rxs1708 are shown in Fig. \ref{rxs_he_spc}. The high-energy emission is consistent with being 100\% pulsed beyond $\sim 150$ keV.

\begin{figure}[t]
\centering
\psfig{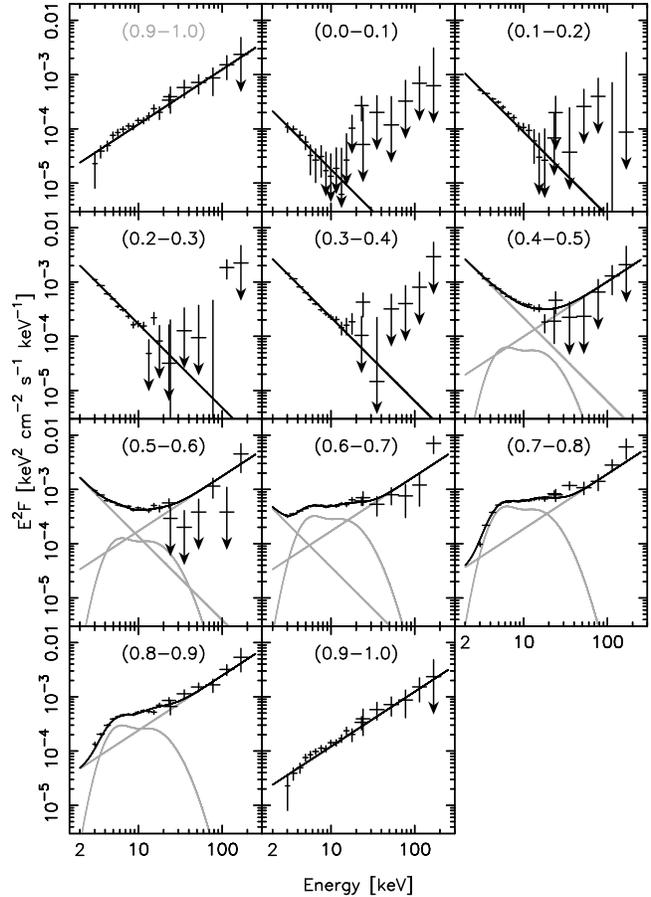}
\caption{Narrow-band ($\Delta\Phi=0.1$) phase resolved spectra of \rxs1708.
Phase intervals are indicated in the panels. Spectral fits using three components are superposed (black) as well as the component
contribution separately (grey). Note the drastic change in shape moving from phase interval 0.9-1.0 to 0.0-0.1: the power-law index switched from very hard, $\Gamma=0.99\pm0.05$, to very soft, $\Gamma=3.58\pm0.34$.}
\label{rxs_phres_spc}
\end{figure}

\begin{figure}[t]
\centering
\psfig{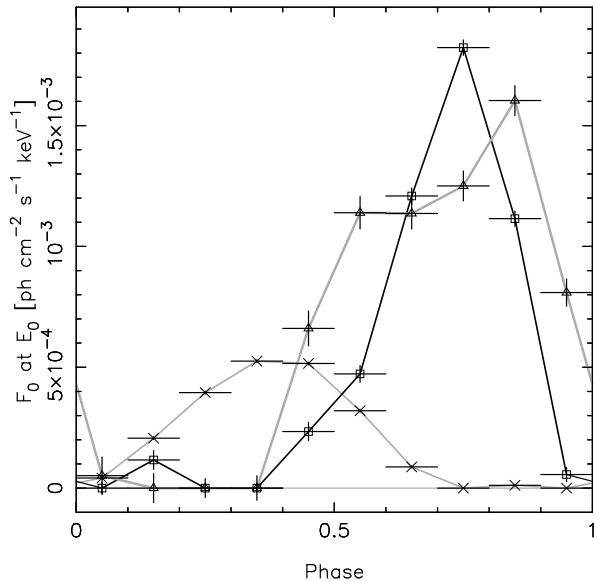}
\caption{Phase distributions of the normalizations, evaluated at 15.04 keV (the pivot energy), of the three spectral-model components. The soft PL component is indicated by crosses and a thin grey line, the hard PL by triangles and a thick grey line, and finally, the composite model by squares and a black line. }
\label{rxs_phdst_nrm}
\end{figure}

The excellent statistics in the timing domain for INTEGRAL ISGRI (20-300 keV) and RXTE PCA (3-30 keV) allowed to explore the spectral
characteristics as a function of phase over the 3-300 keV band. The phase-resolved spectra in ten 0.1 -- wide phase bands over 2 decades in energy are shown as data points in Fig. \ref{rxs_phres_spc}. From these spectra three distinct spectral components with completely different shapes could be recognized: a) a soft PL with index $3.58\pm 0.34$, b) a hard PL with index $0.99\pm0.05$ and c) a composite spectral model contributing over phases 0.7-0.9.
These three spectral components have subsequently been used in spectral fit procedures fitting the data of each phase slice in terms of the sum of three model components, each with a free scale. The phase distributions of the normalizations of these three
components, all evaluated at the pivot energy of 15.04 keV, are shown in Fig. \ref{rxs_phdst_nrm} and represent three decoupled pulse profiles. The soft component peaks around phase 0.4 (c.f. Fig. \ref{rxs_pp}), while the other two components peak around phase
0.8. The width of the curved component ($\sim 0.25$ in phase) is about half the width of the hard component.
These new results from phase resolved spectral analysis give important constraints showing that three dimensional modeling covering both the geometry of the emission regions and the different production processes is required to explain the findings.
 
\subsection{SGR 1806-10}

Persistent high-energy emission of SGR 1806-10 was detected by Mereghetti et al. (2005) and Molkov et al. (2005) using INTEGRAL ISGRI data from 2003-2004 observations, all
taken before the giant flare of December 27, 2004.
They found emission up to $\sim 200$ keV with a power-law shape with photon index in the range 1.5 --1.9, significantly harder than the spectrum of a typical burst showing a thermal bremsstrahlung spectrum. A remarkable property exhibited by this source is that both the spectral hardness and intensity correlate with the burst rate or activity level, which culminated in the giant flare at the end of 2004. 
Pulsed emission, both pre- and post giant flare, is detected up to $\sim 30$ keV using RXTE PCA data. It is {\em not} detected in either RXTE HEXTE or INTEGRAL ISGRI data (Kuiper et al. in prep.), indicating very low values for the pulsed fraction at energies beyond $\sim 20$ keV contrary to what has been observed for the three AXPs showing hard spectral tails.
The morphologies of the pulse profiles show drastic changes as a function of energy, from smooth single peaked (2-5 keV) to double peaked with one (very) sharp pulse in the 10-15 keV band.

\subsection{SGR 1900+14}
G\"otz et al. (2006) reported the detection of hard X-ray emission from SGR 1900+14 above
20 keV using INTEGRAL data collected during 2003-2004 observations. The total 20-100 keV emission spectrum could be described by a power-law model with a rather soft photon index of $3.1\pm 0.5$.
Triggered by this result Esposito et al. (2007) revisited the BeppoSAX observations and found 
in the BeppoSAX observation performed before the August 27, 1998 giant flare indications for a hard spectral tail with a photon index $1.6\pm 0.3$ at a $4\times$ larger flux level in the 20-100 keV band.
This implies a drastic change in spectral behaviour before and after the 1998 giant flare.

%%%%%%%%%%%%%%%%%%%%%%%%%%%%%%%%%%%%%%%%%%%%%%%%%%%%%%%%%%%%%%%%%%%%%%%%%%%%%%%%%%%%%%%%%%%%%%%%

\section{Summary}
From this review it is clear that detailed observations in the high-energy window are crucial for understanding the physical processes taking place under extreme conditions on the surface and in the magnetosphere of magnetars.
A significant fraction of the magnetars currently known emit very luminous  high-energy ($>20$ keV)
radiation. Hard spectral tails have been detected in the total emission from AXPs with power-law indices 
in the range 0.9--1.4, and somewhat softer tails for SGRs. Spectral breaks or bends in the total spectrum of AXPs occur 
above $\sim 250$ keV. The pulsed emission (AXPs) above 10 keV is even harder with indices between -1 and 1. 
For AXPs the pulsed fraction approaches 100\% near 100-150 keV. So far, from SGRs no persistent pulsed emission has 
been detected above $\sim 30$ keV.
We have shown in phase-resolved spectral analyses that distinctly different components contribute to the total pulsed 
emission. These findings require theoretical models involving different production processes taking place at different 
sites in the magnetosphere. 

%%%%%%%%%%%%%%%%%%%%%%%%%%%%%%%%%%%%%%%%%%%%%%%%%%%%%%%%%%%%%%%%%%%%%%%%%%%%%%%%%%%%%%%%%%%%%%%%

\section{MAXI perspective for magnetar research}

With a 10 times larger sensitivity than the RXTE ASM (2-12 keV), observations with the Gas Slit Camera (GSC) of MAXI aboard the International Space Station (ISS) can be very useful for the research on AXPs and SGRs. This instrument sensitive to photons with energies in the range 2-30 keV has 12 cameras with proportional counters totaling a sensitive area of 5350 cm$^2$. 
Due to the scanning nature of the experiment a typical celestial source is twice per ISS orbit for 22 s in the field of view (1\fdeg 5 $\times$ 160\deg) of one of the two camera arrays each consisting of six cameras. This strategy yields a $5\sigma$ sensitivity  of $\sim 1$ mCrab in a one-week accumulation. This is amply sufficient to determine the (total) emission state on a daily base for most of the AXPs which have total fluxes of about 5 mCrab between 2 and 10 keV. Above 10 keV the sensitivity of the MAXI GSC is too low to obtain meaningful results. Due to the sparse covering factor (0.8\% duty cycle) of a source per ISS orbit of $\sim 90$ minutes, there is little chance to catch a fast transient event lasting from seconds to minutes on the fly. Enhanced fluxes levels of AXPs and SGRs lasting from days to weeks can, however, easily be signalled provided that the flux level is sufficiently high.
A point of concern is the rather crude angular response of 1\fdeg5 (FWHM) which will give source confusion in crowded regions near the Galactic plane where the AXPs and SGRs are located.
However, given the time resolution of $120\mu$s of the GSC, for the bright AXPs, \0142, \rxs1708 and \1841, the sensitivity is high enough to extract the pulsed flux, typically 1 mCrab in the 2-10 keV band, from weekly accumulations. This allows also the construction of a pulsar ephemeris, the set of timing parameters determining the rotation behaviour of the pulsar as a function of time. This information is crucial for instruments like INTEGRAL IBIS ISGRI for which the extraction of the pulsed signal relies on pulse-phase folding techniques because of the weakness of the signal.

%%%%%%%%%%%%%%%%%%%%%%%%%%%%%%%%%%%%%%%%%%%%%%%%%%%%%%%%%%%%%%%%%%%%%%%%%%%%%%%%%%%%%%%%%%%%%%%%

\section*{References}

\re
Baring M.G. \& Harding A.K., 2007, Ap\&SS 308, 109

\re
Baring M.G. \& Harding A.K., 2008, AIP Conf. Proc. 968, 93

\re
Belobodorov A.M. \& Thompson C., 2007, ApJ 657, 967

\re
den Hartog P.R., et al., 2004, ATEL \#293

\re
den Hartog P.R., et al., 2006, A\&A, 451, 587

\re
den Hartog P.R., Kuiper L., Hermsen W,  et al., 2008a, A\&A, 489, 245

\re
den Hartog P.R., Kuiper L., Hermsen W, 2008b, A\&A, 489, 263

\re
Esposito P., et al., 2007, A\&A, 461, 605

\re
G\"otz D., et al., 2006, A\&A, 449, L31

\re
G\"otz D., et al., 2007, A\&A, 475, 317

\re
Heyl J.S. \& Hernquist L., 2005a, ApJ 616, 463

\re
Heyl J.S. \& Hernquist L., 2005b, MNRAS 362, 777

\re
Heyl J.S., 2007, Ap\&SS 308, 101

\re
Kaspi V.M, 2007, Ap\&SS 30, 1

\re
Kuiper L., Hermsen W. \& Mendez M. 2004, ApJ 613, 1173

\re
Kuiper L., et al., 2006, ApJ, 645, 556

\re
Mereghetti S., et al., 2005, A\&A 433, L9

\re
Molkov S.V., et al., 2004, Astronomy Letters 30, 534

\re
Molkov S.V., et al., 2005, A\&A 433, L13

\re
Revnivtsev M.G., et al., 2004, Astronomy Letters 30, 382

\re
Thompson C. \& Duncan R.C., 1996, ApJ 473, 322

\re
Thompson C., Lyutikov M., \& Kulkarni S.R., 2002, ApJ 574, 332

\re
Thompson C. \& Belobodorov, 2005, ApJ 634, 565

\re
Woods P. \& Thompson C., 2006, In: Lewin, W., van der Klis,
M. (eds) Compact Stellar X-ray Sources. Cambridge Astrophysics Series,
vol. 39, 547

\label{last}

\end{document}